# Efficient thermionic operation and phonon isolation by a semiconductor-superconductor junction

Emma Mykkänen[1], Janne S. Lehtinen[1], Leif Grönberg[1], Andrey Shchepetov[1], Andrey V. Timofeev[1], David Gunnarsson[1], Antti Kemppinen[1], Antti J. Manninen[1], Mika Prunnila[1,*]

[1]VTT Technical Research Centre of Finland Ltd, P.O. Box 1000, FI-02044 VTT Espoo, Finland.

**Control of heat flux at small length scales is crucial for numerous solid-state devices and systems. In addition to the thermal management of information and communication devices[1] the mastering of heat transfer channels down to the nanoscale also enable, e.g., new memory concepts[2], high sensitivity detectors and sensors[3-5], energy harvesters[6] and compact solid-state refrigerators[7]. Electronic coolers[7,8] and thermal detectors for electromagnetic radiation[5], especially, rely on the maximization of electro-thermal response and blockade of phonon transport. In this work, we propose and demonstrate that efficient electro-thermal operation and phonon transfer blocking can be achieved in a single solid-state thermionic junction. Our experimental demonstration relies on suspended semiconductor-superconductor junctions[8] where the electro-thermal response arises from the superconducting energy gap, and the phonon blocking naturally results from the transmission bottleneck at the junction. We suspend different size degenerately doped silicon chips (up to macroscopic scale) directly from the junctions and cool these by biasing the junctions. The electronic cooling operation characteristics are accompanied by measurement and analysis of the thermal resistance components in the structures indicating the operation principle of phonon blocking in the junctions.**

*mika.prunnila@vtt.fi

Electro-thermal elements are broadly used in various radiation and photon detectors[5], energy harvesters[9] and micro-coolers[8]. The operation of these elements is based on the correlation between heat/energy and particle currents, and the desired operation requires minimization of the thermal conductance to the bath and maximization of the electro-thermal response. Typical examples here are thermoelectric materials[9,10] and thermionic junctions[6,11]. The former are based on diffusive phonon and electron transport along a thermoelectric lead whereas the latter relies on local energy filtering of electrons by an energy barrier. Ultimate physical scaled-down limit of an electro-thermal element is reached by a molecular contact[12,13].

In thermionic junctions the most energetic thermally excited electrons are emitted through vacuum or short solid barriers[11] (Fig. 1a). The physical principle of thermionic junction is therefore quite general and thermionic operation can be observed in electrical and electro-thermal properties of various physical systems, like, pn and Schottky diodes[14], vacuum barrier components[6,15], quantum dots[16], metallic single-electron devices[17], and superconductive tunnel junctions[8,18-21] (Fig. 1b). Clear figure of merit of any electro-thermal element is its ability to cool the cold reservoir below the bath temperature (hot reservoir) by introducing an electron flow (Fig. 1a). Such electronic refrigerator operation has also vast applications in cooling of electronics and different sensors, and Peltier refrigerators based on thermoelectrics are well-known off-the-shelf components dedicated for this task[22].

In all-solid electro-thermal devices, phonons introduce the detrimental heat flow channel that hinders the overall cooling in refrigerators and the total electro-thermal response in harvesters and detectors. Due to this, for example, in thermoelectrics considerable efforts have been devoted in investigations of phonon engineering approaches to reduce phonon heat flow[9,10]. Promising approaches include developing new materials and scaling the cross-sectional dimensions of the thermoelectric leads down to nanometer scale in order to reduce the phonon mean free path. In thermionic junctions connecting different reservoirs (Fig. 1a), thermal isolation has been considered to be difficult to achieve through the junction itself due to apparent strong phonon transmission over short distance. Typical thermal isolation schemes of thermionic devices include utilization of superlattices, vacuum (gas) barriers[6,9], and low electron-phonon coupling occurring at low temperatures[8].

In this Letter, we demonstrate that a single solid interface can operate both as an efficient thermionic element and heat transfer barrier for phonons (Fig. 1a). We use semiconductor-superconductor (Sm-S) tunnel junctions where the electronic thermionic emission is controlled by the superconducting energy gap and voltage bias[19-21]. The phonon thermal boundary resistance ($R_{PTB}$) at the junction provides the phonon blockade (Figs. 1b-d). The Sm-S junctions are used to support, thermally isolate and electronically refrigerate a piece of silicon chip, referred to as the sub-chip (Figs. 1e-g). These suspended junctions provide significant cooling of ~40 % from the bath temperature at sub-kelvin temperatures. The cooler and the overall electro-thermal performance can be significantly enhanced by utilizing phonon engineering methodologies[23,24]. For example, we demonstrate by simulations that such approaches combined with cascaded refrigeration stages can enable cooling even from ~1.5 K to below 100 mK, which is one of the long-standing goals of electronic refrigeration.

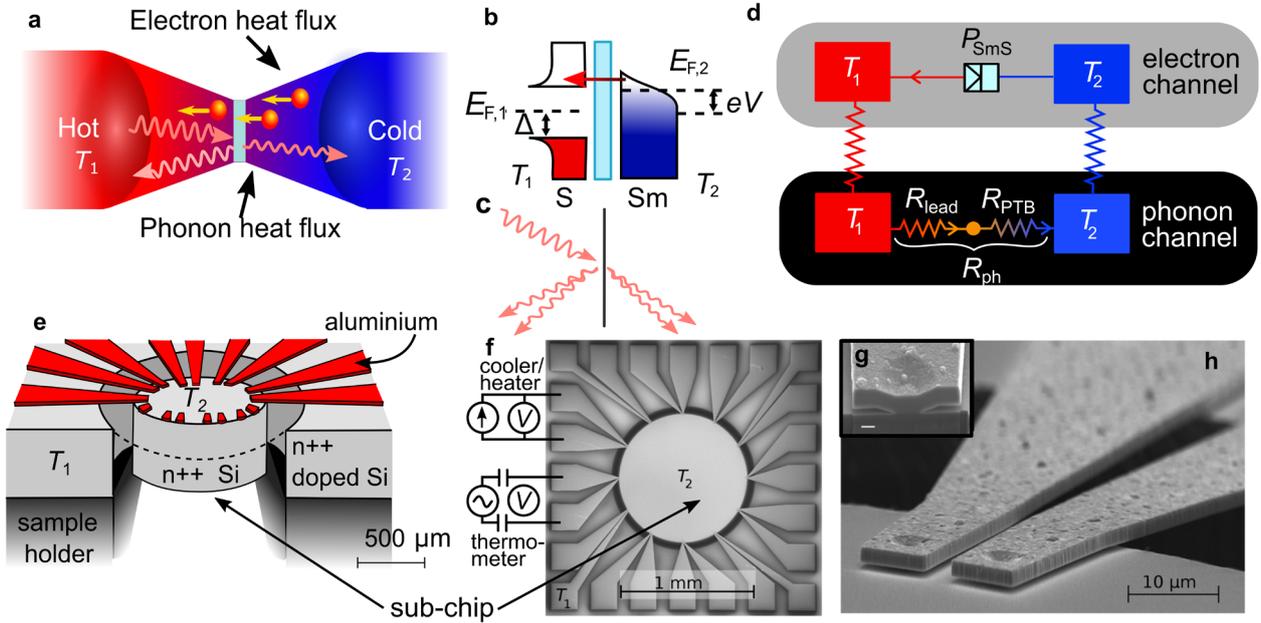

**Figure 1 | Cooling scheme. a,** Conceptual image of particle fluxes through solid-state thermionic barrier/interface connecting hot and cold reservoirs. The interface suppresses phonon heat flow between the reservoirs and acts as a thermionic junction that governs electron transport. **b, c** Electron (**b**) and phonon (**c**) interfaces between the reservoirs. Electron transport is limited by the energy gap of the superconducting lead and phonon transport by $R_{PTB}$. **d,** Simplified heat resistance network of our system presenting the electric heat flow/power ($P_{SmS}$) of the Sm-S junctions and the thermal resistance between the hot and cold reservoirs ($R_{ph}$), which depends on $R_{PTB}$ and lead thermal resistance ($R_{lead}$). For more detailed version see Supplementary figure SI7. **e,** Artistic image of the experimental device. The sub-chip is physically supported by 24 Sm-S junctions connected to the aluminium leads, which mediate the heat conduction between the main chip and sub-chip. The large size of the sub-chip ensures that its electrons and phonons share the same temperature. **f,** Scanning electron micrograph (SEM) of a sample with Ø 1 mm sub-chip. One pair of Sm-S junctions is biased with alternating current (ac) and used as a thermometer whereas one or more of the other pairs are biased with direct current (dc) and used as heaters or coolers. The measurement setup is further discussed in Methods, Supplementary sections 1-2 and Supplementary figure SI1. **g,h,** SEM images of the cross section of a tunnel junction (**g**) and two superconducting leads (**h**), showing that only the Sm-S junctions, with diameters between 1.5 μm and 3 μm, are mechanically connected to the sub-chip.

**Table 1 | Sample parameters and literature values for thermal boundary resistance**. Here $A$ is the area of a tunnel junction, $d$ is the diameter of the sub-chip, $R_A = R_T A$ is the characteristic junction resistance, and $R_T$ is junction resistance. Thermal resistance prefactors ($\alpha$) are the values obtained from the fits of Fig. 2 whereas $\alpha_{\text{AMM}}$ and $\alpha_{\text{DMM}}$ correspond to theoretical $R_{\text{PTB}}$[26]. *Sample S1 had silicon oxide between leads and sub-chip, and therefore the effective area of the phononic heat contact, 137 μm², is larger than the junction area 2.0 μm².

| sample | $A$ μm² | $d$ μm | $R_A = R_T A$ Ωμm² | $\alpha$ K⁴/μW | $\alpha_{\text{AMM}}$ K⁴/μW | $\alpha_{\text{DMM}}$ K⁴/μW | max cooling % | max cooling mK |
|---|---|---|---|---|---|---|---|---|
| S1 | 2.0/137* | 300 | 538 | 4.5 | - | - | 15 (@ 140 mK) | 22 (@ 166 mK) |
| S2 | 7.5 | 1000 | 476 | 5.6 | 6.5 | 8.8 | 40 (@ 170 mK) | 83 (@ 244 mK) |
| S3 | 3.2 | 300 | 1945 | 29 | 16 | 21 | 29 (@ 173 mK) | 56 (@ 220 mK) |

The fabrication of our Sm-S (Si-Al) junctions follows the procedures of Ref. 25. The sub-chip is released from the main chip using deep reactive ion etching and HF vapour etching (see Methods for details). Scanning electron micrographs of one of the devices are shown in Figs. 1f-h. In this Letter we show data on three samples (S1-3) with sample parameters as given in Table 1. In contrast to the others, sample S1 had silicon oxide between the aluminium leads and the sub-chip, which decreased $R_{\text{PTB}}$ and thus yielded information on the thermal resistance of the Al leads.

An Sm-S (or normal metal - insulator - superconductor, NIS) junction can function as thermionic element due to the energy filtering property of the superconductor, which originates from the superconducting energy gap $\Delta$ of the quasiparticle (electron) density of states (figure 1b).[19,20] At small bias voltages ($V \leq \Delta/e$, where $e$ is elementary charge) only the most energetic electrons of the sub-chip can overcome the gap and tunnel to the superconductor, which is equivalent to cooling. A similar effect is obtained at the opposite bias when quasiparticles tunnel to the lowest unoccupied states of the sub-chip. Here we use a simple model for the Sm-S junction cooling power (for more rigorous approach, see supplementary section 4)

$$P_{\text{SmS,cool}}(T_2) \approx \frac{\Delta^2}{e^2 R_T} 0.59 \left(\frac{k_B T_2}{\Delta}\right)^{\frac{3}{2}} - \frac{1}{2}\frac{V_{\text{opt}}^2}{R_{\text{gap}}}, \quad (1)$$

where $R_T$ is tunneling resistance and $T_2$ the temperature of the sub-chip. The first term in (1) is the ideal cooling power, which is valid at optimal voltage[27], $V_{\text{opt}} \approx (\Delta - 0.66 k_B T_2)/e$, when $T_2 \leq T_S \ll \Delta/k_B$.[20] Here $k_B$ is the Boltzmann constant and $T_S$ is the temperature of the superconductor. The latter term is an approximate model for heating due to the sub-gap leakage resistance $R_{\text{gap}} = R_T/\gamma$ of nonideal tunnel junctions ($\gamma$ is a leakage parameter), which can originate, e.g., from quasiparticle states within $\Delta$[28] or traps and dopants in Sm-S junctions[25].

The electronic refrigeration results with the suspended Sm-S junctions fabricated directly on the sub-chip are presented in Fig. 2a. Direct refrigeration of both electrons and phonons of the macroscopic sub-chip are enabled by large degenerately doped sub-chip and the phonon transport and transmission bottlenecks between the sub-chip and the main chip. The former assures negligible thermal resistance between electrons and phonons and the latter brings in thermal isolation. Note that this is in strong contrast to previous works, in which the NIS junctions on the substrate at bath temperature refrigerated only the electrons of the normal metal, not phonons, and a cold finger was used for indirect refrigeration of the payload[29-31]. Our best performing sample is S2 (sub-chip with 1 mm diameter and 0.4 mm height). Its maximal absolute and relative temperature reductions are 83 mK (at 244 mK) and 40 % (at 170 mK), respectively. The cooling power for this sample is about 2 pW/μm² at 300 mK, which is of similar magnitude as those achieved with the most efficient NIS junctions[32]. The calculated curves of Fig. 2a are obtained as the balance between cooling power (1) and thermal resistance results of Fig. 2b (see below), and they fit the experimental data well when the sub-gap leakage resistance is taken into account.

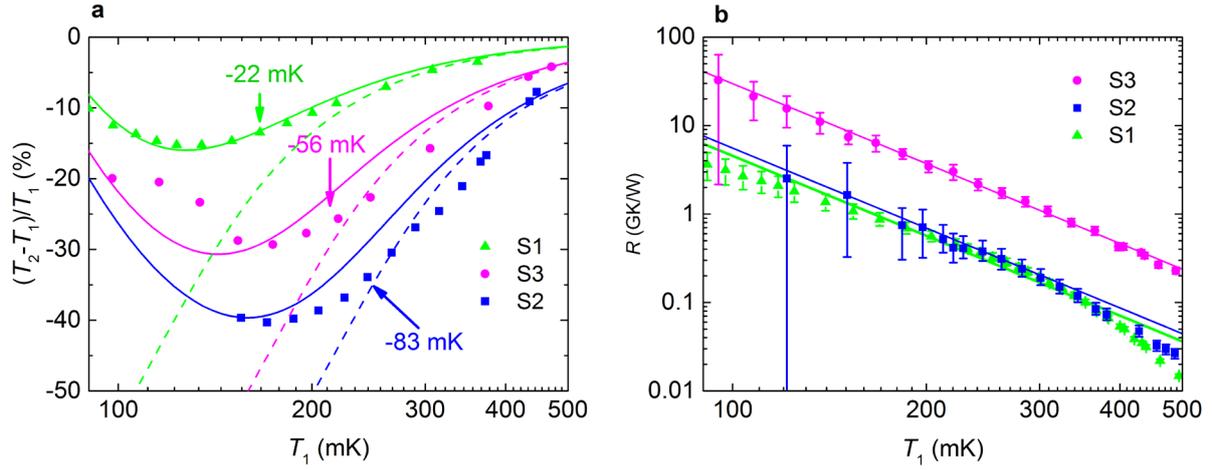

**Figure 2 | Measurement results. a,** Relative temperature difference between the sub-chip and main chip as a function of bath temperature for samples S1, S2 and S3 at optimal cooling voltage. The largest absolute temperature reduction is indicated by arrows. Refrigeration is modelled by setting the heat leak ($T^{-3}$ fits of b) in balance with the cooling power produced by ideal Sm-S junctions (dashed curves) or by Sm-S junctions that have finite sub-gap leakage (solid curves). **b,** Thermal resistance between the sub-chip and its environment for samples S1, S2 and S3, and $T^{-3}$ fits (solid lines). Error margins are dominated by the uncertainty of simulation parameters and by measurement uncertainty (see Supplementary section 8 for details).

The detrimental heat leak from the main chip to the sub-chip can be represented by the total thermal resistance $R$, the measurement results of which are shown in Fig. 2b. The heat leak is mainly via the superconducting leads and tunnel junctions. Quasiparticle backflow from the superconductor is also significant in some cases (see Supplementary Sections 3, 4 and 8). Electronic heat leakage through the tunnel junction is effectively blocked by the superconducting energy gap and the tunnelling resistance. In addition, the thermal coupling between the electron and phonon systems in the superconductor is suppressed at low temperatures. Therefore, we view the system as two (phononic) thermal resistances in series: the Al-Si interface and the aluminium lead, as shown in Fig. 1d.

The thermal resistances, $R$, of Fig. 2b were measured by biasing a set of the cooler junctions above $\Delta$, which results in heating, $P_{\text{SmS,heat}}$. Then $R = \delta T/P_{\text{SmS,heat}}$, where $\delta T = T_2 - T_1$ is the measured temperature difference (see Methods and Supplementary section 5). The experimental data follows closely the fitted lines with $T^{-3}$ dependence, which is expected for both the thermal boundary resistance of the Al-Si interface[26] and the thermal resistance arising from polycrystalline aluminium wires[33].

Table 1 compares the fitted prefactors, $\alpha$, of thermal resistance $R(T) = \alpha T^{-3}$ to the expected thermal boundary resistance ($R_{\text{PTB}}$) based on acoustic mismatch (AMM) and diffusive mismatch (DMM) models[26]. The former utilizes ballistic transmission of acoustic waves and the latter diffusive scattering at the interface. The phonon thermal resistance of the aluminium leads, $R_{\text{lead}}$, can be of the same order of magnitude as $R_{\text{PTB}}$, but it depends on phonon mean free path, which is not well-known. However, $R_{\text{PTB}}$ is proportional to the interface area between the sub-chip and leads whereas the wire thermal resistance is the same in all samples. We interpret the results as follows (for detailed analysis and discussion, see Supplementary Sections 5-7): (*i*) Because of the large interface area (see Table I), the thermal resistance of sample S1 constitutes mainly from that of the leads. (*ii*) The thermal resistance of sample S3 is dominated by the phononic interfacial thermal resistance resulting in that $R$ is close to AMM and DMM values. (*iii*) The thermal resistance of S2 is close to both that of S1 and the theories for $R_{\text{PTB}}$ and thus is likely to have contributions from both $R_{\text{PTB}}$ and $R_{\text{lead}}$. We cannot rule out near-field heat transfer effects[34] either: S2 has the largest surface area and the thermal photon wavelength exceeds the sub-chip to main-chip distance by several orders of magnitude. (*iv*) At higher temperatures, thermal resistances decrease below the fits for samples S1 and S2. This originates most likely from the increased electron-phonon coupling in the leads, which creates an additional heat conduction channel (see Supplementary section 6).

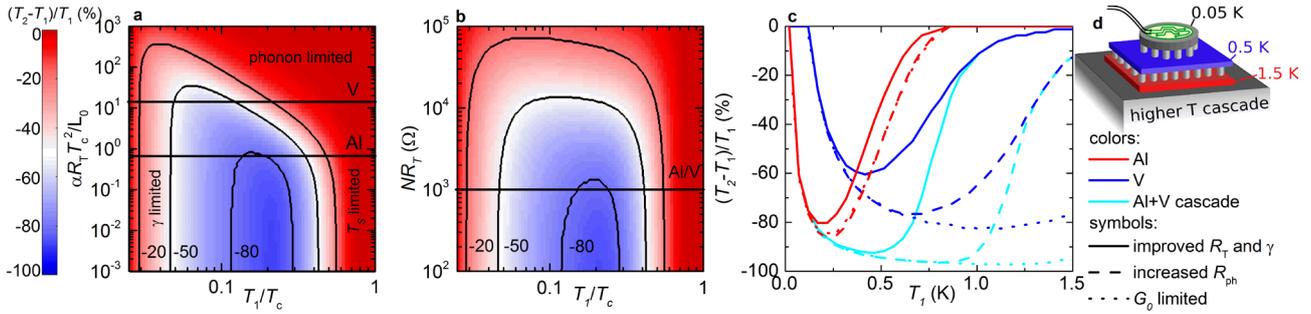

**Figure 3 | Future prospects. a,b,c** The achievable relative temperature reduction with any superconductor with critical temperature $T_c$ when $\gamma = 10^{-3}$. The horizontal lines in a and b and data of c are calculated with $R_A =100\ \Omega\mu m^2$. **a** Cooling against $R_{PTB}$. The y-axis is normalized with Lorentz number, $L_0$. Black horizontal lines show values for aluminium ($T_{c,Al}$= 1.184 K) and vanadium ($T_{c,V}$= 5.38 K) based junctions, when parameters of S3 are used for $R_{ph}$, and correspond to the solid lines in c. **b** Constriction limited cooling as a function of $NR_T$, where $N$ is the number of thermal conductance quanta in the constriction. The black horizontal line corresponds to the dotted lines in c, and is calculated when $N$ equals 10 quanta per tunnel resistance of $R_T =100\ \Omega$. **c** Cooling with improved tunnel junction parameters, size dependent quantum effects and cascaded coolers. The red lines present data for aluminium, blue for vanadium and cyan for a cascaded cooler, where the sample is first cooled with vanadium-silicon junctions and then with aluminium-silicon junctions. The solid lines indicate a case, where the cooler performance is improved but the thermal resistance to the environment is the same as in sample S3. The dashed lines present a situation, where this resistance is increased by 10 compared to its original value, and the dotted lines present the temperature reduction when both $R_{PTB}$ and the phonon resistance of the constriction are taken into account. **d** Artistic image of a cascaded cooler.

The cooling power of the refrigerator can be increased to allow larger payloads simply by increasing the number of tunnel junctions. However, reduction of achievable minimum temperature requires the increase of cooling power compared to the heat leak from environment. In Fig. 3, we analyse the benefits of improved tunnel junction quality, utilization of nanoscale effects of phonon heat conduction (phonon engineering), and cascaded superconductors with different $\Delta$[35,36].

The cooler performance can be improved by decreasing sub-gap leakage or junction resistance as seen from equation (1). Our coolers had $\gamma \approx 4\ldots 5.5 \times 10^{-3}$, but $\gamma \sim 10^{-4}$ has been achieved in Ref. 25. Furthermore, we estimate that the characteristic junction resistance can be decreased to about 100 $\Omega\mu m^2$. Figure 3a shows that using these parameters and the thermal resistance of sample S3, maximum relative refrigeration of about 80 % can be achieved with Al-Si junctions in narrow temperature range. Importantly, such junctions are already close to the regime where the refrigeration is limited by $\gamma$ at low temperatures and mainly by $T_S$ at high temperatures instead of the phonon transport, which means that $R_{PTB}$ is a sufficient thermal barrier for Al-based refrigeration.

The applicable cooling regime depends on the critical temperature of the superconductor, $T_c \approx \Delta/1.764 k_B$. A vanadium ($T_{c,V}$=5.4 K) based cooler has been experimentally demonstrated to yield improved performance over aluminium ($T_{c,Al}$=1.2 K) at higher temperatures[37]. However, the cooling power follows $P \propto \Delta^2 f(T/T_c)$, where $f$ is a function independent of $\Delta$ (see Eq. (1)), whereas the heat leaks follow $\dot{Q} \propto \alpha T^4 \propto \alpha \Delta^4 (T/T_c)^4$ (see Eq. (2) in Methods), which have a different $\Delta$ dependence when $T$ is scaled with $T_c$. Therefore, in Fig. 3a, V-Si junctions are in the regime where refrigeration would benefit from improved thermal isolation.

Nanoscale effects of phonon heat conduction are actively studied for, e.g., improved thermal barriers, higher efficiency of energy harvesting, and thermal management of nanoscale electronics[23,24]. Our concept can significantly benefit from phonon engineering methodology, where phonon transport is further suppressed, e.g., at the sub-chip by using a superlattice close to the tunnel junctions or at the tunnel junction interface by weakening chemical bonding[23]. However, the most straightforward and practical approach is to constrict the superconducting lead by multiple nanowires[38]. When the diameter of the nanowire is reduced, the phonon mean free path and thermal resistivity scale with the diameter, but the electrical resistivity remains constant[39]. In this diffusive limit, increasing $R_{ph}$ by factor of 20 would move the vanadium line in Fig. 3a to the regime of sufficient phonon heat blockade. Improving the thermal resistance is ultimately limited by the conductance (or resistance) quantum $G_0 = 1/R_0 = \pi^2 k_B^2 T/3h$, where $h$ is Planck constant[40,41]. Figure 3b shows simulations of cooling in this limit where heat leak follows $\dot{Q} \propto G_0 T \propto \Delta^2(T/T_c)^2$ and has the same $\Delta$-dependence as the cooling power.

Figure 3c collates the simulation data on improved junctions, improved thermal resistance, and a cascaded cooler with both V and Al based junctions. We envision that the cascaded cooler should follow the scheme shown in Fig. 3d, which is closer to the one used in commercial thermoelectric coolers than in typical SINIS cascade coolers[35]. Our simulations demonstrate the effects of two levels of improved $R_{\text{ph}}$: when it is increased by factor 10 and when it reaches the ultimate thermal conductance quantum limit. As a result, V-based refrigeration is dramatically improved whereas an Al-based device is only slightly affected. These simple models omit quasiparticle effects, but Supplementary section 9 shows that temperature reduction of about 50% can be achieved up to about 1.5 K also when quasiparticles are taken into account.

In summary, we have proposed and demonstrated efficient thermionic operation and phonon transfer blocking in solid state junctions. We utilized Si-Al semiconductor-superconductor junctions, where the thermionic operation arises from the superconductive gap, and the phonon transfer blocking naturally occurs due to phonon thermal boundary resistance. Ultimate figure of merit of thermionic junction is its ability to cool a thermal mass below the bath temperature and we demonstrated relative cooling of 40 % of a mm-scale suspended silicon chip. We suggested, supported by simulations, phonon engineering methods to reach cooling from above 1 K to sub-1K temperatures, which can enable replacing cryo-liquid based refrigeration stages by solid state ones[7,8]. Phonon isolated thermionic junctions could be also used in thermal energy harvesting and in different thermal photodetectors (bolometers), which have vast applications from chemical sensing to security[3,5,42]. They would be particularly useful in the latter, where the phonon noise arising from thermal conductance to the bath is one of the fundamental limitations of sensitivity. Molecular contacts exhibit strong phonon isolation[12,13] and these could be used in realizing the ultimate scaled-down limit of the thermionic concepts discussed in this Letter.

# Methods

### Sample fabrication
We used 0.4 mm thick highly doped silicon wafers to assure negligible thermal resistance between electrons and phonons in the sub-chip. First the surface doping was further increased with phosphorous implantation to reduce the Schottky barrier between Al and Si. Then a 476 nm of tetraethyl orthosilicate (TEOS) was deposited onto the wafer by low pressure chemical vapour deposition (LPCVD) and the wafers were annealed to densify the oxide and to activate the implanted dose. The oxide was patterned with UV-lithography and the tunnel junctions were prepared as described in Ref. 25, which was directly followed by sputter deposition of 1000 nm thick aluminium film. The electrodes were patterned by UV-lithography and plasma etching. The sub-chip was released with Bosch etch process using silicon oxide as stopping layer. Finally, the excess silicon oxide was removed with hydrogen fluoride (HF) vapour.

### Measurement setup
The samples were cooled in a dilution refrigerator with the minimum temperature of about 20 mK. Each sample had in total 24 Sm-S junctions. One pair of junctions was used as a thermometer and 1 to 11 pairs as coolers. Due to the limited number of cables in the cryostat, some Sm-S junctions were biased using a common cable. Since the resistance of the cabling between sample and measurement equipment was large (around 1240 Ω per cable) compared to the junction tunneling resistance (from 50 Ω to 620 Ω), the cooler voltages were determined by 4-probe measurements when necessary. Unused junctions were left floating. The schematics of the measurement setup is presented in Supplementary figure SI1.

### Thermometry
Since Sm-S junction is sensitive to temperature, its current-voltage characteristics can be used for thermometry. In the measurement, a pair of Sm-S junctions was capacitively coupled to the measurement setup and biased with alternating current ($f = 21.11$ Hz). The resulting ac voltage was measured with a voltmeter. The signal was then Fourier transformed and the peak at drive frequency recorded. Each Sm-S thermometer was calibrated against a standard ruthenium oxide thermometer, which in turn had been calibrated against Coulomb blockage thermometer[43]. The measurement configurations are further discussed in Supplementary section 1 and Supplementary figure SI1.

### Cooling measurements
With samples S1 and S3, 22 out of 24 Sm-S junctions were used as coolers and the remaining pair as thermometer. With S2 we used only 20 junctions for cooling. When possible, cooler junctions were biased as Sm-S-Sm pairs in order to keep the sub-chip at virtual ground.

The optimal bias voltage for cooling depends on temperature. Therefore, each cooler needs to be adjusted both when bath temperature is changed but also after additional cooling is provided by the other cooler junctions (see Supplementary figure SI4 for detail). Due to the limited number of cryostat lines, not all Sm-S-Sm junction pairs could be biased individually but some of them were used as sets of 1 to 3 pairs. Because of this and variation in sample parameters, the optimum voltage of the voltage source was different depending on the cooler.

The optimization scheme for multiple coolers was iterative for S1 and S2: First, the optimum voltage was measured for each cooler while others were at zero voltage. Then the procedure was repeated multiple times but now the inactive coolers were biased to voltages, which had given the maximum temperature reduction in the previous step. The optimization scheme for S3 was otherwise similar as for S1 and S2, but the sample was biased over each line instead of pairs of lines and the sub-chip was not at virtual ground.

**Thermal resistance measurements**

The total thermal resistance between sub-chip and its environment was measured by heating the sub-chip electrically with Sm-S junctions and recording its electron temperature increase. For samples S1, S2, and S3, the number of junctions used for heating was 2, 4, and 6, respectively. Most heat currents, $\dot{Q}$, originating from a single process, follow

$$\dot{Q} = \frac{1}{\alpha n}(T_1^n - T_2^n), \tag{2}$$

where $n$ is a constant. Adding more processes can be done by summation. When temperature differences are small this simplifies to $\dot{Q} = \frac{1}{\alpha} T^{n-1} \delta T$, which gives the thermal resistance $R = \frac{\delta T}{\dot{Q}} = \alpha T^{-n+1}$. In equilibrium, the same amount of heat is inserted and removed from the sub-chip and thus $\dot{Q}$ is equal to the heating or cooling power of Sm-S junctions, which can be written as[20]

$$P_{\text{SmS}}(V, T_2, T_S) = \frac{\Delta^2}{e^2 R_T} \int_{-\infty}^{\infty} d\epsilon (\epsilon - eV/\Delta) n(\epsilon, \gamma)[f(\epsilon, T_S) - f(\epsilon - eV/\Delta, T_2)]. \tag{3}$$

Here $f(\epsilon, T_i) = [1 + \exp(\epsilon \Delta/k_B T_i)]^{-1}$ is the Fermi function at temperature $T_i$ and $n(\epsilon, \gamma) = \text{Re}\left[(\epsilon + i\gamma)/\sqrt{(\epsilon + i\gamma)^2 - 1}\right]$ is the density of states where non-zero leakage parameter $\gamma$ describes non-idealities in the junctions[28]. The method described in Supplementary section 3 was used to determine $T_S$, and other parameters of Eq. (3) were obtained from the current-voltage characteristics (Supplementary section 2). Thermal resistance is calculated as a linear fit to the $P_{\text{SmS}}$ vs. $\Delta T$ data. In the analysis, we used only data where $|eV| \geq 2\Delta$ and $\Delta T \leq 20$ mK…40 mK. An example plot of the measurement results and further discussion on thermal resistance analysis are presented in Supplementary figure SI5 and section 5.

**Methods only references**

**Acknowledgement**

The authors thank Jukka Pekola and Ilari Maasilta for useful discussions and Kestutis Grigoras for taking cross-section images of the Al-leads. This project was financially supported by H2020 programme FET-open project EFINED (project number 766853), H2020 programme project MOS-QUITO (project number 688539) and Academy of Finland projects QuMOS (project numbers 288907, 287768). E.M. acknowledges support from Jenny and Antti Wihuri foundation. This work was performed as part of the Academy of Finland Centre of Excellence program (project 312294).